\documentclass[11pt,twocolumn]{article}
\usepackage{epsfig}
\usepackage{graphicx}
\begin{document}

\renewcommand{\thefootnote}{\fnsymbol{footnote}}

\twocolumn[\begin{center} {\LARGE\bf{Anomalous Stark Shifts in
Single Vertically Coupled Pairs of InGaAs Quantum Dots}}\\
\vspace{0.5cm}
\bf{\underline{R. Oulton}$^{1,3,\ast}$,
A.I.Tartakovskii$^{1}$, A. Ebbens$^{1}$, J.J. Finley$^{4}$, D.J.
Mowbray$^{1}$, M.S.~Skolnick$^{2}$ and M. Hopkinson$^{2}$}\\
\small \it{$^{1}$Dept. of Physics and Astronomy, University of Sheffield,
Sheffield, UK, S3 7RH}\\\it{$^{2}$Dept. of Electrical and
Electronic Engineering, University of Sheffield, Sheffield, UK, S3
7RH}
\it{$^{3}$Experimentelle Physik II,
Universit\"at Dortmund, D-44221 Dortmund, Germany}\\
\it{$^{4}$Walter Schottky Institut, Technische Universi\"at M\"unchen, D-85748, Garching, Germany,}\\
\it{$^{\ast}$Corresponding author: phone +49 (0)231 755 3536; fax
+49 (0)231 755 3674;\\ e-mail oulton@e2a.physik.uni-dortmund.de}
\end{center}

\vspace{0.5cm} {\bf{Abstract:}} \small{Vertically coupled Stranski
Krastanow QDs are predicted to exhibit strong tunnelling
interactions that lead to the formation of hybridised states.  We
report the results of investigations into single pairs of coupled
QDs in the presence of an electric field that is able to bring
individual carrier levels into resonance and to investigate the
Stark shift properties of the excitons present.  Pronounced
changes in the Stark shift behaviour of exciton features are
identified and attributed to the significant redistribution of the
carrier wavefunctions as resonance between two QDs is achieved. At
low electric fields coherent tunnelling between the two QD ground
states is identified from the change in sign of the permanent
dipole moment and dramatic increase of the electron
polarisability, and at higher electric fields a distortion of the
Stark shift is attributed to a coherent tunnelling effect between
the ground state of the upper QD and the excited state of the
lower QD.} \vspace{0.5cm}


\vspace{0.5cm}\vspace{1.0cm}

]The ideal quantum dot has the extremely useful property that excitons contained within it have an extremely long coherence time at low temperature, comparable to its radiative lifetime of hundreds of picoseconds.[1]  Several schemes for quantum computation have been proposed which exploit this long coherence time, the most promising of which appear to be those where the excitons in a pair (or eventually series) of quantum dots form an entangled state via a tunnelling interaction between them.[2,3]  Stranski-Krastanow grown QDs are ideal for this purpose, as the strain field resulting from the formation of one dot layer facilitates the growth of a second dot layer just a few nm above the first, with an almost 100$\%$ probability that a second QD forms directly above the first.[4]\\
\begin{figure}[h]
\centerline{\epsfxsize=8cm\epsfbox[0 20 550 780]{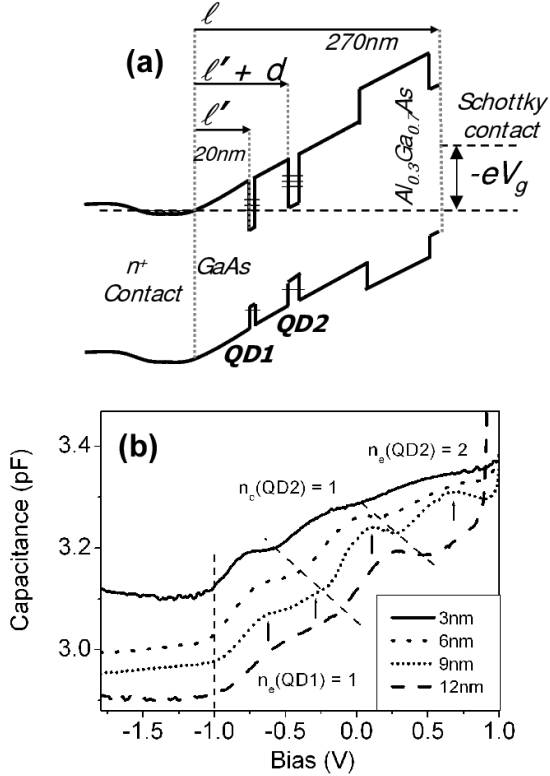}}
\caption{\small{(a)Schematic of bandstructure under applied
reverse bias $V_g$. $l$, $l'$ and $l'+d$ give distances from
n-type contact to top contact, lower QD1 and upper QD2
respectively. (b)Capacitance-voltage traces from four coupled
layer samples, with interdot distance, d = 3,6,9,12nm.  Negative
bias values denote reverse bias.  Dashed lines indicated onset of
charging.}}
\end{figure}

\indent We report the results of optical spectroscopic measurements of single pairs of vertically coupled Stranski-Krastanow quantum dots in the presence of an external electric field.  The electric field allows the carrier levels in each QD to be brought into resonance with one another, and also gives access to the exciton Stark shift, allowing valuable insight into the dramatic carrier redistribution that occurs in the QDs as coherent tunnelling occurs.\\
\indent Fig. 1(a) shows a schematic of the sample structure investigated.  The sample was grown using molecular beam epitaxy on an undoped GaAs substrate, and has a structure as follows: 1$\mu$m undoped (UD) GaAs, 500 nm n+ (n=5x$10^{18} cm^{-3}$) GaAs contact layer, 20 nm UD tunnel barrier, two QD layers, then 105nm UD GaAs, 75nm UD Al$_{0.3}$Ga$_{0.7}$As, and finally a 5nm GaAs cap. The wafer rotation was stopped before the QD layer was grown ($\sim6$ ML of In$_{0.5}$Ga$_{0.5}$As deposited at 530 ${^o}$C) to produce a QD density gradient across the wafer. Wafer rotation was resumed, a 12nm UD spacer layer of GaAs was then grown to separate the two QD layers, and again stopped for the second QD layer, grown with an identical deposition rate and direction to the first.  After growth, a low-density region was chosen, Ohmic contacts were established to the n+ contact layer and a 200 nm thick Au shadow mask was then deposited on the Schottky contact into which 800 nm diameter microapertures were opened lithographically.  Photoluminescence spectra were taken using a conventional $\mu PL$ setup with excitation performed at low intensity at an energy of $\sim1380meV$.  Several QDs are accessed underneath each aperture.\\
\indent The sample design is such that the electron and hole states of the upper dot (QD2) show a stronger shift in energy with bias than the lower dot (QD1) allowing the carrier levels of the upper dot to be tuned through those of the lower dot.  The energy shift of the levels is determined by the lever arm [5] of each QD level, and is given by
\begin{equation}
E_1^{e,h} = e(V_g-\frac{V_{GaAs}}{2})\frac{l'}{l}+E_{0}
\label{eq1}
\end{equation}
\begin{equation}
E_2^{e,h} =e(V_g-\frac{V_{GaAs}}{2})\frac{l'+d}{l}
\label{eq2}
\end{equation}\\
where  $E_1^{e,h}$ and  $E_2^{e,h}$ are the energy shifts of the carrier levels in the lower and upper dots respectively, $V_g$ is the applied reverse bias, $(eV_{GaAs}/2)$ is the half the bandgap of GaAs (0.75eV), {\it l} is the distance between the n-type layer and the sample surface, {\it l'} is the distance between the n-type layer and the lower QD, 20 nm in this case, and d is the nominal distance between the QD layers, 12 nm for the spectra discussed here.  $E_0$ is the difference in energy (offset energy) between the lower and upper carrier levels at zero electric field (flatband).  The relationship between the offset energy $E_0$ and reverse bias at which resonance occurs, $V_0^{e,h}$, may be easily extracted from Eqs. (1) and (2) to give
\begin{equation}
 E_0 =e(V_0^{e,h}-0.75V) {d\over{l}}
\label{eq3}
\end{equation}\\
\indent Tuning the bias aligns the carrier levels of the upper and lower QDs.  At this point, the wavefunction of each level has a finite probability of occupying the other, and the two QD states become hybridised.  In this case a level of hybridisation, and therefore a coherent tunnelling interaction, will be observed for a wide range of detuning from the single QD resonance.\\
\begin{figure}[h]
 \centering
\centerline{\epsfxsize=9cm\epsfbox[75 250 550 553]{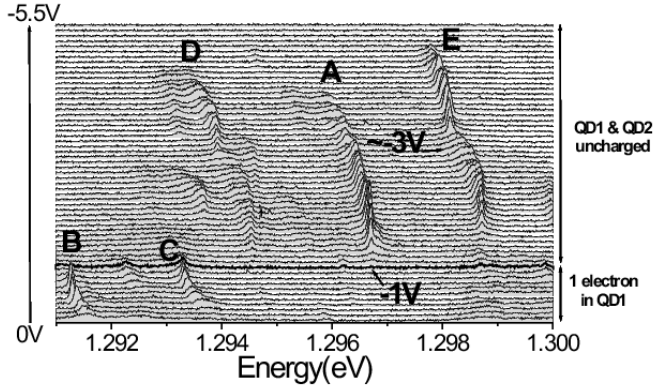}}
\caption{\small{PL spectra of coupled layer of QDs separated by
12nm, with reverse bias increasing from 0V to -5.5V.  The thick
line spectrum corresponds to -1V, the bias at which QD1 becomes
charged with 1 electron.  Above -1V both quantum dots remain
uncharged from the contact.}}
\end{figure}
\indent Capacitance-voltage ($C-V_g$) measurements of four similar coupled layer samples are shown in Fig. 1b.  The lower QD layer has a constant separation from the doped layer of 20nm.  The upper layer is separated from the lower layer by the length d, with values d = 3, 6, 9 and 12 nm.  The $C-V_g$ traces clearly show several features that arise due to charging from the contact, similar to that observed by refs [5,6,7].  Charging occurs first for the lower QD at $\sim -1V$ reverse bias in all cases.  The bias at which charging of the upper dot occurs is dependent on the QD spacing $d$.  The second and third prominent peaks that are observed are clearly dependent on the QD spacing, and therefore may be attributed to charging of the upper QD.  For the 12nm sample we therefore conclude that the upper QD becomes charged from the contact at $\sim$0V.  Fig. 2 shows a series of PL spectra taken over a bias range $0.0 \leq V_g \leq 5.5 V$.  Several features (A, D, E) are observed at high bias.  At $\sim -1V$ all of the features quench, and a new set of features appears between $-1.0 \leq V_g \leq 0.0 V$ (B, C).  This energy shift of the features, arising due to the change in the Coulomb interaction, is in good agreement with the $C-V_g$ data that show that electron charging to the lower QD occurs at this bias range.\\
\indent All of the features observed in Fig 2 exhibit a shift in energy as the reverse bias is increased.  In single QDs the energy shift with increasing electric field is well-understood, and for moderate electric fields is given by the Quantum Confined Stark Effect, namely:
\begin{equation}
 E =E_0 +pF +\beta F^2
\label{eq4}
\end{equation}\\
where E is the ground state exciton emission energy, F is the electric field parallel to the growth direction, $E_0$ is the ground state energy at $F = 0$, $p$ is the vertical permanent dipole moment, and $\beta$ is the vertical exciton polarisability.  The permanent dipole moment is a measure of the vertical electron hole separation, $r$, in the QD, and is given by $p = e.r$ where e is the electronic charge.  $\beta$ is a measure of the rate of separation of the electron and hole with field, and is expressed as the sum of the electron and hole polarisabilities  $\beta_e + \beta_h$.[9]\\
\indent Several of the features observed in Fig. 2 show a quadratic energy dependence on electric field, for example features A, B, C.  A curious property of these features is that as the lower QD becomes charged (i.e. going from features A,D,E, to B,C), a change in sign of the polarisability is observed.  The features B and C have a minimum energy and Stark shift to higher energy as the bias is lowered to -1V (positive curvature), whereas for features A, D, E, and for single QDs grown with similar conditions,[8] a Stark shift to lower energy is observed as the bias is lowered (negative curvature).\\
\indent The permanent dipole moment of the features also change sign.  This is apparent when considering the maximum/minimum points of the features highlighted.  The turning point in energy occurs at a field $F_0$, given by $F_0 = p/2\beta$  from Eq.(4).  The feature A has a turning point at negative $F_0$ (reverse bias) due to the fact that $p$ is positive (i.e. electron above hole) and $\beta$ is negative.  However, the features B and C also have turning points at negative $F_0$. Given that $\beta$ is positive, this implies that $p$, the permanent dipole moment, is now negative (electron below hole).  We make the assumption that the features B, C arise from the same QDs as the features A,D,E, as it is highly unlikely that two separate QDs are observed over different bias ranges.  Thus the change in sign of the dipole must be due to a change in the exciton configuration in the same QD.  A positive polarisability suggests that as the field is increased, one of the carrier levels increases in energy.  For a "square-well-like" structure this is unphysical.  Moreover, such a positive polarisability has never been observed for single QDs.  The change in sign of both $p$ and $\beta$ strongly suggests that a dramatic change in the wavefunction takes place.  The effects would be difficult to explain using a picture of two decoupled states, and therefore a more plausible explanation appears to be that these effects are the result of a tunnelling coupling of the two QDs.  The centre of mass of the electron wavefunction is now between the two QDs, and with the hole in the upper QD this results in a negative dipole (we demonstrate later that the features A,D,E, and hence also B and C, must arise from the upper QD).  Note that excitons may also be created in the lower QD also.  However, because the hole level energies will be different for the two QDs, an electron from the lower QD will not appear in this detection range.  If the upper exciton remained quantum mechanically decoupled from the "spectator" electron in the lower QD, only weak perturbative Coulomb interactions would apply: we note that changes in polarisability have not been observed for any charged exciton species.\\
\indent We suggest therefore that as the bias is lowered, the electron ground states of the two QDs come into resonance.  The coherent tunnelling appears to occur at the point where the lower QD becomes charged with an electron from the contact.  This sharp transition to the coherent tunnelling regime is a result of the Coulomb blockade effect that occurs when a QD becomes charged with a single electron.  The Coulomb blockade has the effect of increasing the energy of the lower QD electron level by several $meV$.  An estimate of this value was made by Finley $et. al.$[8] for similar QDs, and was found to be $\sim 25 meV$.\\
\begin{figure}[h]
 \centering
\centerline{\epsfxsize=9cm\epsfbox[67 352 550 619]{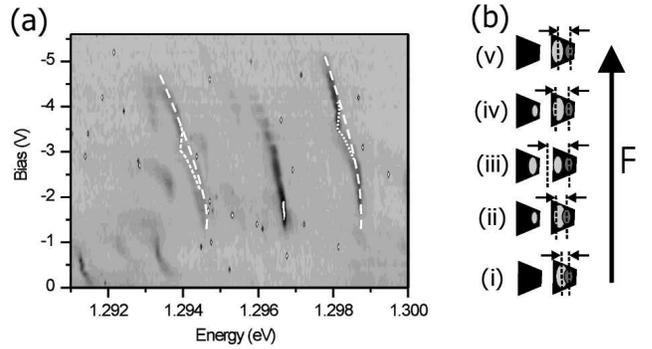}}
\caption{\small{(a)Grey-scale plot of spectra shown in Fig. 2.
Dark areas indicate regions of high intensity.  Dashed lines
indicate quadratic Stark shift of features D,E labelled in Fig 2.
Dotted lines indicate the anomaly in Stark shift in reverse bias
range $-2.5 \leq V_g \leq -3.5$. (b)Suggested configuration of
electron (light grey) and hole (dark grey) as upper electron level
is tuned from non resonant (i and v) to resonant (iii) with the
lower QD electron level.  Vertical lines indicate the centre of
mass of the electron and hole wavefunction.}}
\end{figure}
\indent At $V_g = -1V$ we make the assumption that the ground state electron level in the upper QD is higher in energy than that of the neutral electron level in the lower QD.  Thus no coupling occurs between the two levels.  However, as the QD becomes charged, the energy level of the ground state in the lower QD undergoes a discrete increase of $25 meV$.  The electron levels of the two ground states are now close in energy, and the coupling strength is strong enough that a coherent tunnelling coupling occurs for the bias range $-1.0 \leq V_g \leq 0.0 V$.\\
\indent The features that appear at higher bias (A,D,E) appear to show a conventional (single QD-like) behaviour over most of the bias range, with $e.g.$ feature A having a positive dipole (electron above hole at zero field) of the order of $\sim 4 \AA$ and negative polarisability.  This is expected: at higher bias the ground state levels are not close to resonance and single QD-like behaviour is observed.  However, at the specific bias range $-2.5 \leq V_g \leq -3.5V$ the features D and E show a distinctive anomaly in the Stark shift.  Fig 3(a) shows a grayscale intensity plot of the features observed in Fig 2.  As a guide to the eye, the dashed line in the figure shows the Stark shift at low and high fields, and the dotted line shows the deviation from this Stark shift at $-2.5 \leq V_g \leq -3.5V$ for both features D and E.\\
\indent We interpret this kink in the features D and E as a coherent tunnelling of a carrier between the upper and lower QDs, similar to that observed at lower bias, occuring in this case when a carrier of the ground state exciton becomes resonant with an excited state.  The bias range over which we observe the deviation is small ($\sim 1V$ corresponding to detuning energy of the order $20-30meV$ from Eqs.1,2), suggesting that we observe a weak tunnelling effect only.\\
\indent The bias at which the resonance occurs, $-3V$, allows one to calculate the energy difference between the two QD states at zero field using Eq.(3) to be $E_0 \sim 140meV$.  The resonance between the two electron ground states has already been identified at lower bias.  Given that previous studies of similar single QDs have estimated the hole ionisation energy to the wetting layer to be $\sim 90meV$, a resonance between two hole states seems unlikely.  The most likely explanation therefore is that we observe a resonance between the electron ground state in the upper dot, and an excited electron state in the lower dot.  This explanation is supported by the fact that the feature becomes increasingly broad as it comes into resonance.  The linewidth of the features at low field is $<100\mu eV$ whereas at $\sim3V$ a linewidth of $\sim 500\mu eV$ is observed.  In previous work it was shown that the excited states of Stranski-Krastanow QDs are subject to fast dephasing processes of the order of a few ps.  It appears likely therefore that an electron ground state hybridised with such an excited state will also be subject to such fast dephasing mechanisms.\\
\indent As the ground state comes into resonance with the excited state, a redistribution of the wavefunction results in a significant alteration of the values $p$ and $\beta$, to a first approximation constant for a single QD.  It is this change in $p$ and $\beta$ that leads to the "kink" observed.  Fig. 3(b) shows a suggested schematic of the electron-hole distribution of the two quantum dots as a function of field, depicting the constant surface of the electron occupation probability in light grey and of the hole in dark grey, that would explain the behaviour observed.  We consider the situation where the electron level of the exciton in the upper QD comes into resonance with an electron level in the lower QD at $V_g \sim 3V$.  As the two levels approach resonance, the electron wavefunction in the upper QD begins to penetrate into the lower QD.  This results in the centre of mass of the wavefunction moving significantly towards the lower QD.  Even a weak penetration into the lower QD will lead to significant changes in the permanent dipole moment of the exciton.  The maximum deviation from the quadratic Stark shift at $\sim -3V$ corresponds to the point where the levels are exactly in resonance and the coherent tunnelling is maximum.\\
\indent In conclusion, we have studied anomalous Stark shift effects that occur as two electron levels of vertically coupled QDs are brought into resonance with each other using an electric field.  As coherent tunnelling processes occur, the wavefunction is significantly altered, a fact that is apparent in fundamental changes of the permanent dipole moment and polarisability.  These changes may also be explained by considering a coupled QD system, and therefore allow us to identify resonances between two ground states, and between a ground and an excited state of two QDs.

\section*{References}
\small{
$[1]$ Borri, P., W. Langbein, S. Schneider, U. Woggon, R.L. Sellin, D. Ouyang, D. Bimberg {\it Phys. Rev. Lett.} {\bf 87}, 157401 (2001)
\\$[2]$ Bayer, M., P. Hawrylak, K. Hinzer, S. Fafard, M. Korkusinski, Z. R. Wasilewski, O. Stern, A. Forchel {\it Science} {\bf 219}, 451 (2001)
\\$[3]$ Szafran, B., S. Bednarek and  J. Adamowski {\it Phys. Rev. B} {\bf 64}, 125301 (2001)
\\$[4]$ Xie, Q., A. Madhukar, P. Chen, N. Kobayashi {\it Phys. Rev. Lett.} {\bf 75}, 2542 (1995))
\\$[5]$ Medeiros-Ribeiro, G., F.G. Pikus, P.M. Petroff, A.L. Efros, {\it Phys. Rev. B} {\bf 55}, 1568 (1997)
\\$[6]$ Warburton, R.J., C. Schäflein, D. Haft, F. Bickel, A. Lorke, K. Karrai, J.M. Garcia, W. Schoenfeld, P.M. Petroff, {\it Nature} {\bf 405}, 926 (2000)
\\$[7]$ Finley, J.J., P.W. Fry, A.D. Ashmore, A. Lema\^itre, A.I. Tartakovskii, R. Oulton, D.J. Mowbray, M.S. Skolnick, M. Hopkinson, P.D. Buckle, P.A. Maksym {\it Phys. Rev. B} {\bf 63}, 161305(R) (2001)
\\$[8]$ Finley, J.J., M. Sabathil, P. Vogl, G. Abstreiter, R. Oulton, A.I. Tartakovskii, D.J. Mowbray, M.S. Skolnick, S.-L. Liew, A.G. Cullis, M. Hopkinson {\it submitted to Phys. Rev. B.} (2004)
\\$[9]$ Barker, J.A., E.P. O'Reilly, {\it Phys. Rev. B.} {\bf 61}, 13840 (2000)}
\end{document}